\documentclass{amsart}

\usepackage{amssymb,amsthm,euscript,amsmath,graphicx}

\newtheorem{rem}{Remark}[section]

\def\static{_{{\rm s}}} 
\def\rev{^{{\rm r}}}  
\def\drho{\partial_\rho} 
\def\visc{^\text{v}}    
\def\fl#1{_\text{#1}}
\def\Euler{\fl{Euler}}  
\def\SchM{\fl{SchM}}    
\def\Lan{\fl{Lan}}      
\def\Alt{\fl{Alt}}      

\begin{document}
\title{Weakly nonlocal fluid mechanics - the Schr\"odinger equation}
\author{P. V\'an, T. F\"ul\"op}
\address{Budapest University of Technology and Economics\\
Department of Chemical Physics\\
1521 Budapest, Budafoki \'ut 8.}
\email{vpet@phyndi.fke.bme.hu}

\date{\today}

\begin{abstract}
A weakly nonlocal extension of ideal fluid dynamics is derived
from the Second Law of thermodynamics. It is proved that in the
reversible limit the additional pressure term can be derived from
a potential. The requirement of the additivity of the specific
entropy function determines the quantum potential uniquely. The
relation to other known derivations of Schr\"odinger equation
(stochastic, Fisher information, exact uncertainty) is clarified.
\end{abstract}

\maketitle


\section{Introduction}

Weakly nonlocal or coarse grained or gradient are attributes of
theories from different fields of physics indicating that in
contradistinction the traditional treatments, the governing
equations of the theory depend on higher order gradients of the
state variables. Weakly nonlocal is a nomination in continuum
physics dealing with internal structures \cite{Mar02a}, coarse
grained appears in statistically motivated thermodynamics
\cite{HohHal77a}, and gradient is frequently used in mechanics
\cite{Bed00a,CimKos97a}). The simplest way to find weakly nonlocal
equations can be exemplified by the Ginzburg-Landau equation which
can be considered as a first weakly nonlocal extension of a
homogeneous relaxation equation of an internal variable. The
traditional derivation of the Ginzburg-Landau equation is based on
a characteristic mixing of variational and thermodynamic
considerations. One applies a variational principle for the static
part and the functional derivatives are introduced as
thermodynamic forces into a relaxation type equation. A clear
variational derivation to obtain a first order differential
equation is impossible without any further ado (e.g. without
introducing new variables to avoid the first order time
derivative, which is not a symmetric operator) \cite{VanMus95a}.
One can apply these kinds of arguments in continuum theories in
general, preserving the doubled theoretical framework separating a
reversible and an irreversible parts of the equations
\cite{GrmOtt97a,OttGrm97a}. However, there are also other
systematic attempts to unify the two parts with different
additional hypotheses \cite{Gur96a,Mar02a} and to eliminate this
inconsistency of the traditional approach.

The ultimate aim is to find a unified and predictive theoretical
framework that involves higher order gradients in the governing
equations of physics. All the approaches mentioned above are
systematic in the sense that the gradient dependent terms are not
introduced in an ad-hoc way, the Second Law of thermodynamics
plays an important role in each of them. An analysis of the
involved thermodynamics shows that in case of the Ginzburg-Landau
equation one can determine the variational part purely from the
requirement of a nonnegative entropy production, without any
additional assumptions \cite{Van02m2}. This way of thinking shows
why and in what sense and under what conditions is the
Ginzburg-Landau form is distinguished among the possible other
weakly nonlocal equations of a single internal variable (called
order parameter in special cases).

However, the Ginzburg-Landau equation is only a prototypical
example and several different weakly nonlocal differential
equations can appear in continuum physics. It is well-known that
quantum mechanics is a nonlocal theory. However, nonlocality is
understood in different ways in the different treatments. A clear
interpretation and appearance of nonlocality in quantum mechanics
is that the Bohmian quantum potential depends on the derivatives
of the quantum probability density \cite{Boh51b,PenCet96b}. In
this sense, quantum mechanics is a special weakly nonlocal (or
coarse grained or gradient) fluid mechanics.

In this paper we investigate all possible weakly nonlocal
extensions of fluid mechanics that are compatible with the Second
Law. A straightforward application of Liu's procedure gives the
surprising consequence that a weakly nonlocal extension of
traditional fluid mechanics in the density leads to a
generalization of the Euler equation that incorporates the
hydrodynamic model of quantum mechanics. Moreover, we will show
that this model is distinguished among the different possible
weakly nonlocal fluids in the sense that the characteristic Fisher
information like form of the gradient dependent part of the
entropy density function is a unique consequence of its isotropic
and additivity properties. After these considerations we treat
shortly some interpretational and conceptual problems, too. We
argue that our investigations lead to a derivation of quantum
mechanics that is preferable according to the principle of Occam's
razor because it is based on a minimal number of assumptions.

The starting point of our investigations is fluid mechanics.
However, one should not consider it as a phenomenological theory
with a rich and definite microscopic-molecular background, but
rather as an empty bottle of general physical principles that are
valid in any continuum (field) theory (e.g. balance of momentum,
conservation of mass, Second Law, ...). The approach of
non-equilibrium thermodynamics fills this bottle with content
using a step-by-step approach. Therefore, arriving at the
so-called hydrodynamic model of quantum mechanics does not prevent
us from transforming it to a wave function form and enjoying the
advantages of a linear equation. However, we should know the
physical content of the linearity and be aware of the price. For
example it is well-known that the Schr\"odinger equation is not a
reference frame independent, objective equation. Moreover, it
cannot be written in a frame independent form because of its
energetic origin \cite{Mat86b}. Similarly, canonical quantization
is also frame/observer dependent (see e.g. \cite{Bol98a}).
However, the theories based on a momentum balance (e.g. the
hydrodynamic model) are frame independent and objective
\cite{FulKat98a}.

\section{Fluid mechanics in general}

The {\em basic state space} of one-component fluid mechanics is
spanned by the density $\rho$ and the velocity ${\bf v}$ of the
fluid. Hydrodynamics is based on the balance of mass and the
balance of momentum \cite{Gya70b}. Classical fluid mechanics is
the theory where the constitutive space, the domain of the
constitutive functions, is spanned by the basic state space
$(\rho,{\bf v})$ and the gradient of the velocity $\nabla {\bf
v}$. The pressure/stress tensor is the only constitutive quantity
in the theory. The pressure function defines the material in
continuum physics. For example, it determines whether it is a
fluid or a solid, and whether it is a material in local
equilibrium without memory or not, etc. Introducing higher order
gradients of the basic variables into the constitutive space one
can obtain weakly nonlocal extensions regarding the density and
the velocity. In this paper we investigate the weakly nonlocal
extension of classical fluid dynamics in the density.

The balance of mass in local substantial form can be written as
\begin{equation}
  \dot{\rho}+\rho\nabla\cdot {\bf v} = \sigma_m,
\label{cond_mass}\end{equation}

\noindent where $\rho$ is the density, ${\bf v}$ is the velocity,
$\sigma_m$ is mass production, and the dot above the quantities denotes
the substantial (comoving) derivative.

The balance of momentum, i.e. the Cauchy equation, is
\begin{equation}
\rho\dot{\bf v}+\nabla\cdot {\bf P} = \rho {\bf f},
\label{cond_mom}\end{equation}

\noindent where {\bf P} is the pressure and {\bf f} is the force
density. For discussing the characteristics of continuum materials
the effects given by the source terms do not play any role. The
Second Law requires that the production of the entropy is
nonnegative in insulated and source-free systems. Particularly in
our case if there is no production of mass ($\sigma_m =0 $) and
there are no external forces ($\rho{\bf f} = {\bf 0}$), then the
production of entropy must be nonnegative
\begin{equation*}
\rho\dot{s}+\nabla\cdot {\bf j}_s \geq 0.
\label{aim_mech}\end{equation*}

The constitutive quantities in the above equations are the
pressure {\bf P}, the specific entropy $s$, and the conductive
current ${\bf j}_s$ of the entropy. According to the Coleman-Mizel
form of the Second Law, the entropy inequality is a constitutive
requirement \cite{MusEhr96a}, that is, we are looking for
constitutive functions that solve the inequality. Thus the entropy
inequality should be a pure material property, it is required to
be valid independently of the initial conditions. In the
exploitation of the inequality Liu's theorem plays an important
technical role. Details of the different state spaces, more
detailed description of thermodynamic concepts, and the applied
mathematical methods (especially the Liu procedure) can be found
in \cite{MusAta01a}; regarding the weakly nonlocal extension see
\cite{Van03a,Van02m2}.

Let us remark that, in fluid systems, the balances of different
(kinetic, potential, internal and total) energies should all be
considered \cite{Ver97b}. However, in our treatment, dealing with
nonlocal extensions only in the density, the internal energy
changes do not play role, therefore we avoided the computational
complications coming from additional governing equations. In case
of moving fluids it is better to regard the function $s$ as a
generalized (kinetic) potential \cite{GlaPri71b}. However,
avoiding the unnecessary extension of the terminology and
emphasizing the very thermodynamic point of view of the
argumentation, we call it entropy.

It is important that, in case of moving continua, the substantial
time derivatives appear in the treatment, because these time
derivatives express the time changes of the corresponding physical
quantities in a frame fixed to the moving continuum. These time
changes are independent of any external reference frames as one
can see clearly in a frame independent treatment of continuum
mechanics \cite{Mat84a,Mat93b}.

\section{Nonlocal fluid mechanics - the Schr\"odinger-Madelung
\\equation}

Fluid mechanics is treated in the case when the {\em basic state
space} is that of classical hydrodynamics ($\rho, {\bf v}$) and
the {\em constitutive space} contains gradients of the density
$\rho$ in addition to the classical case and is spanned by the
variables $(\rho, \nabla\rho, {\bf v}, \nabla{\bf v}, \nabla^2\rho
)$. Here $\nabla^2\rho$ denotes the second space derivative of
$\rho$ (sometimes written as $\nabla \circ \nabla \rho$, where
$\circ$ is the traditional notation of the tensorial/dyadic
product in hydrodynamics). The {\em space of independent
variables} is spanned by the next time and space derivatives of
the constitutive variables, that is, by $(\dot{\rho},
\nabla\dot{\rho}, \dot{\bf v}, \nabla\dot{\bf v},
\nabla^2\dot{\rho}$, $\nabla^2{\bf v}, \nabla^3\rho)$, as a
consequence of the entropy inequality.

In applying the Liu procedure, the constraints are the balance of
mass (\ref{cond_mass}) and the balance of momentum
(\ref{cond_mom}). Moreover, because of the higher derivatives of
density in the constitutive space, the space derivative of the
mass balance is also to be considered as a constraint,
\begin{equation}
\nabla\dot{\rho}+\nabla\rho\nabla\cdot {\bf v} + \rho\nabla\nabla \cdot
{\bf v} = 0.
\label{cond_massder}\end{equation}

\noindent This situation is similar to what happen in case of the
thermodynamic derivations of the Ginzburg-Landau equation
\cite{Van02m2}, or in relativistic constitutive theories
\cite{HerAta98p}. On the other hand, higher order space
derivatives of equation (\ref{cond_massder}) cannot be
constraints, because those contain derivatives that have already
been included in the space of independent variables.

Now we apply Liu procedure, with the method of Lagrange-Farkas
multipliers (see \cite{Van02m2,Liu72a}) as
\begin{eqnarray*}
\rho D_1s \dot{\rho} &+&
\rho D_2s \cdot(\nabla\rho\dot)  +
\rho D_3s \cdot \dot{\bf v} +
\rho D_4s :(\nabla {\bf v}\dot) +
\rho D_5s :(\nabla^2\rho\dot) + \\
D_1 {\bf j}_s \cdot\nabla\rho &+&
D_2 {\bf j}_s :\nabla^2\rho +
D_3 {\bf j}_s :\nabla{\bf v} +
D_4 {\bf j}_s \cdot:\nabla^2{\bf v} +
D_5 {\bf j}_s \cdot:\nabla^3\rho + \\
\Gamma_1 (\dot{\rho} &+& \rho\nabla\cdot {\bf v}) +
\Gamma_2 \cdot(\nabla\dot{\rho} + \nabla\rho\nabla\cdot {\bf v} +
    \rho\nabla\nabla \cdot {\bf v}) +
\Gamma_3 \cdot(\rho\dot{\bf v} + \nabla\cdot {\bf P}) \geq 0.
\end{eqnarray*}

\noindent Here the multipliers $\Gamma_1, \Gamma_2$, and
$\Gamma_3$ were introduced for the constraints (\ref{cond_mass}),
(\ref{cond_mom}) and (\ref{cond_massder}), respectively. The
subscript numbers denote derivation according to the corresponding
variable in the constitutive space, $(\rho, \nabla\rho, {\bf v},
\nabla{\bf v}, \nabla^2\rho)$, e.g. $D_1s = \frac{\partial
s}{\partial\rho}$. The source terms in the balances have been
considered as zero. For what follows, it is important to observe
that the substantial time derivative does not commute with the
space derivative (gradient), instead, the following identities are
to be applied:
\begin{eqnarray*}
(\nabla a\dot) &=& \nabla\dot{a} -
    \nabla {\bf v}\cdot \nabla a, \\
(\nabla^2 a\dot) &=& \nabla^2 \dot{a} -
    \nabla^2 {\bf v}\cdot \nabla a  -
    2 \nabla {\bf v}\cdot \nabla^2 a.
\end{eqnarray*}

\noindent For the sake of easier applicability we give these equations
with indices, too:
\begin{eqnarray*}
(\partial_i a\dot) &=& \partial_i \dot{a} -
    \partial_i v_j\partial_j a, \\
(\partial_i \partial_j a\dot) &=& \partial_i \partial_j \dot{a} -
    \partial_i \partial_j v_k \partial_k a  -
    2 \partial_i v_k \partial_k \partial_j a .
\end{eqnarray*}

Here $i,j,k = 1,2,3$ denote the Cartesian coordinates. Using these
identities the terms in the above inequality can be rearranged as
\begin{eqnarray*}
(\rho D_1s - \Gamma_1)\dot{\rho} &+&
(\rho D_2s - \Gamma_2)\cdot\nabla\dot{\rho} +
\rho(D_3s - \Gamma_3)\cdot\dot{\bf v} +
\rho D_4s : \nabla\dot{\bf v} + \\
\rho D_5s : \nabla^2\dot{\rho} &+&
(D_5{\bf j}_s - \Gamma_3\cdot D_5{\bf P})\cdot : \nabla^3\rho +
(D_4{\bf j}_s - \Gamma_3\cdot D_4{\bf P})\cdot :
    \nabla^2{\bf v} - \\
\rho\Gamma_2 \cdot \nabla\nabla\cdot{\bf v} &-&
\rho D_5s:(\nabla^2{\bf v}\cdot\nabla\rho +
    2 \nabla{\bf v}\cdot\nabla^2\rho)-
\rho D_4s:(\nabla{\bf v}\cdot\nabla{\bf v}) + \\
(D_1{\bf j}_s - \rho D_2s\cdot\nabla{\bf v} &-&
    \Gamma_2\nabla\cdot{\bf v} -
    \Gamma_3\cdot D_1{\bf P})\cdot\nabla\rho +
(D_2{\bf j}_s - \Gamma_3\cdot D_2{\bf P}):\nabla^2\rho + \\
(D_3{\bf j}_s - \rho\Gamma_1 {\bf I} &-&
    \Gamma_3\cdot D_3{\bf P}):\nabla{\bf v} \geq 0
\end{eqnarray*}

The multipliers of the independent variables are the Liu
equations, respectively:
\begin{eqnarray}
\rho D_1s &=& \Gamma_1, \label{L1}\\
\rho D_2s &=& \Gamma_2, \label{L2}\\
D_3 s &=& \Gamma_3, \label{L3}\\
D_4s &=& {\bf 0}, \label{L4}\\
D_5s &=& {\bf 0}, \label{L5}\\
(D_4 j_s)_{jkl} &-& (\Gamma_3)_i(D_4 P)_{ijkl}
    - (\Gamma_2)_l \rho (\delta_{il}\delta_{jk} +
    \delta_{jl}\delta_{ik})/2 = 0. \label{L7}\\
(D_5{j}_s)_{jkl} &-& (\Gamma_3)_i(\partial_3P)_{ijkl}
    = 0 \label{L6}
\end{eqnarray}

\noindent Here, the last two equations are given with indices to
avoid misunderstanding. Equation (\ref{L6}) is symmetric in every
tensorial component (for all permutations of $j$, $k$ and $l$),
and (\ref{L7}) is symmetric in $j$ and $k$ because of the symmetry
of the corresponding independent variables. In the following we
are to solve Liu's equations (\ref{L1})-(\ref{L7}). As a
consequence of (\ref{L4}) and (\ref{L5}), the specific entropy
does not depend on the second gradient of $\rho$ and on the
gradient of ${\bf v}$. Hence, $s(\rho,\nabla\rho, {\bf
v},\nabla{\bf v},\nabla^2\rho) =s(\rho,\nabla\rho,{\bf v})$.
(\ref{L1})-(\ref{L3}) give the Lagrange-Farkas multipliers as
derivatives of the entropy. Therefore, from a thermodynamic point
of view, they are a kind of generalized intensive variables in the
theory \cite{KirHut02p}. Let us treat the entropy as a primary
physical quantity, that is we want to express the other
constitutive functions with its help. Now, one can give a solution
of (\ref{L6}) and (\ref{L7}) as
\begin{eqnarray*}
{\bf j}_s(\rho,\nabla\rho, {\bf v},\nabla{\bf v},\nabla^2\rho) &=&
    \Gamma_3\cdot{\bf P}
    + \frac{1}{2}\rho\Gamma_2 \cdot \nabla{\bf v}
    + {\bf j}_0(\rho, \nabla\rho, {\bf v}) =\\
 &=& D_3 s\cdot{\bf P}
    + \frac{1}{2}\rho^2\left(D_2s\nabla\cdot{\bf v}
    + D_2s\cdot\nabla{\bf v}\right)
    + {\bf j}_0(\rho,\nabla\rho, {\bf v}),
\end{eqnarray*}

\noindent where ${\bf j}_0$ is an arbitrary (differentiable)
function. Thus Liu's equations can be solved and yield the
Lagrange-Farkas multipliers as well as restrictions for the
entropy and the entropy current. Applying these solutions of the
Liu equations, the dissipation inequality can be written as
\begin{eqnarray*}
\rho\dot{s} &+& \nabla\cdot{\bf j}_s =\\
&=& \nabla\cdot{\bf j}_0
    + (\nabla D_3 s):{\bf P}
    + \left[(\frac{\rho^2}{2}(\nabla\cdot D_2s {\bf I}
        + \nabla D_2s)
    -\rho^2 D_1s {\bf I}\right]:
    \nabla{\bf v} \geq 0.
\end{eqnarray*}

Here ${\bf I}$ denotes the second order unit tensor ($\delta_{ij}$).
Let us now define a {\em traditional fluid} with a specific entropy
of the following form
\begin{equation}
    s(\rho,\nabla\rho,{\bf v}) =
        s\static(\rho,\nabla\rho) - \frac{{\bf v}^2}{2}.
 \label{tradf_s}\end{equation}

\noindent From this form it is clear that $s\static$ corresponds
to a static (equilibrium) specific entropy and $s$ is a kind of
general non-equilibrium entropy closely connected to the kinetic
potential of Glansdorff and Prigogine \cite{GlaPri71b}, as it was
mentioned in the introduction. However, here we exploited the
entropic representation of the variables and $s\static$ depends
also on the gradient of the density function. Now it is reasonable
to introduce a new notation of the derivatives as $\partial_\rho
:= D_1$ and $\partial_{\nabla\rho} := D_2$.

Assuming an {\em ordinary entropy current} with ${\bf j}_0={\bf
0}$ we can write the dissipation inequality and the entropy
current as
\begin{equation}
-\nabla {\bf v} :\left({\bf P}
    - \left[\rho^2\left( \frac{1}{2}\nabla\cdot\partial_{\nabla\rho}
        s\static
    - \partial_\rho s\static\right){\bf I}
    -\frac{\rho^2}{2}\nabla\partial_{\nabla\rho} s\static
        \right]\right) \geq 0,
\label{tradf_dineq}\end{equation}

\begin{equation}
{\bf j}_s = -{\bf v}\cdot{\bf P}
    + \frac{\rho^2}{2}\left(\partial_{\nabla\rho}s_s\nabla\cdot{\bf v}
    + \partial_{\nabla\rho}s\static\cdot\nabla{\bf v}\right).
\label{entcur_gen}\end{equation}

The advantage of the entropy (\ref{tradf_s}) of a traditional
fluid is that the inequality (\ref{tradf_dineq}) is solvable,
because it has the force-current form of irreversible
thermodynamics and contains the pressure as a  single dynamic
constitutive function ($s\static$ is static and assumed to be
known). We can define {\em nonlocal reversible pressure} as
\begin{equation}
{\bf P}\rev(\rho,\nabla\rho) :=
    \frac{\rho^2}{2}\left[\left(
        \nabla\cdot\partial_2s\static - 2 \partial_1s\static\right){\bf I}
        + \nabla\partial_2s\static
    \right].
\label{P_rev}\end{equation}

\noindent If the total pressure is of this form then the entropy
production is zero, there is no dissipation, the theory is
reversible (conservative). If the entropy is local (independent of
the gradient of the density) then we obtain
\begin{equation}
{\bf P}\rev\Euler(\rho) := \rho^2 \drho s\static(\rho) {\bf I},
\end{equation}

\noindent therefore, the corresponding equations are of the ideal
Euler fluid, where $p(\rho) = \rho^2 \drho s\static(\rho)$ is the
scalar pressure function. From this form we can identify the
relation of our generalized potential and the traditional entropy.

Introducing the viscous pressure ${\bf P}\visc$ as usual, we can
solve the dissipation inequality (\ref{tradf_dineq}) and give the
corresponding Onsagerian conductivity equation as
$$
{\bf P}\visc := {\bf P} - {\bf P}\rev = {\bf L}\cdot \nabla{\bf
v}.
$$

\noindent Here {\bf L} is a nonnegative constitutive function. Let
us recognize that if $s\static$ is independent of the gradient of
the density, {\bf L} is constant, and ${\bf P}\visc$ is an
isotropic function of only $\nabla{\bf v}$, then we obtain the
traditional Navier-Stokes fluid.

One can prove easily that the reversible part of the pressure of a
traditional fluid is {\em potentializable}, {\it i.e.}, there is a
 scalar valued function $U$ such that
\begin{equation}
    \nabla\cdot {\bf P}\rev = \rho\nabla U.
\label{potcond}\end{equation}

\noindent $U$ can be calculated from the entropy function as
\begin{equation}
    U = \nabla\cdot(\rho\partial_{\nabla\rho}s\static) -
    \partial_\rho(\rho s\static).
\label{qpres_gen}\end{equation}

Therefore in case of reversible fluids the momentum balance can be
written alternatively as
\begin{equation}
\rho\dot{\bf v} + \nabla\cdot{\bf P}^r = {\bf 0} \quad
\Longleftrightarrow \quad \dot{\bf v} + \nabla U = {\bf 0}.
\label{balNew}\end{equation}

Giving the form of $s\static$, we obtain some specific nonlocal
fluids.

\underline{Schr\"odinger-Madelung fluid}. Here the entropy is
defined as
\begin{equation}
    s\SchM(\rho,\nabla\rho,{\bf v}) =
    -\frac{\nu\SchM}{2}\left(\frac{\nabla\rho}{2\rho}\right)^2
    - \frac{{\bf v}^2}{2} =
    -\frac{\nu\SchM}{8} (\nabla \ln\rho)^2 - \frac{{\bf v}^2}{2}.
\label{Schf_s}\end{equation}

\noindent where $\nu\SchM$ is a constant scalar. The corresponding
reversible pressure is
\begin{equation}
{\bf P}\rev\SchM =
    - \frac{\nu\SchM}{8}\left(\Delta\rho {\bf I} + \nabla^2\rho
    - \frac{2 \nabla\rho\circ\nabla\rho}{\rho}\right),
\label{P_Sch}\end{equation}

\noindent where $\circ$ denotes the tensorial/dyadic product, as
mentioned before. The potential is
\begin{equation}
U\SchM = - \frac{\nu\SchM}{4\rho}\left(\Delta \rho -
        \frac{(\nabla\rho)^2}{2\rho}\right)
    = -\frac{\nu\SchM}{2} \frac{\Delta R}{R},
\label{U_B}\end{equation}

\noindent where we introduced $R=\sqrt{\rho}$ to show more clearly
that (\ref{U_B}) is the quantum potential in the de Broglie-Bohm
version of quantum mechanics (if $\nu\SchM = \hbar^2/m^2$)
\cite{Boh51b,Hol93b}.

Further, the entropy current of the Schr\"odinger-Madelung fluid is
\begin{equation}
    {\bf j}\SchM = -{\bf v}\cdot{\bf P}\rev\SchM
       - \frac{\nu\SchM}{8}(\nabla\rho\nabla\cdot{\bf v}
       + \nabla\rho\cdot\nabla{\bf v}).
\label{js_Sch}\end{equation}

\begin{rem} In the hydrodynamic model of quantum mechanics the
pressure cannot be determined uniquely, because one concludes from
the Sch\"odinger equation that a neutral quantum fluid preserves
the vorticity. Therefore in the hydrodynamic model the pressure is
calculated from the potentializability condition (\ref{potcond})
(from the potential) and thus one can add a curl of any function
of the density and its gradient without changing the physics
(Bernoulli equation). For example, replacing $\Delta\rho {\bf I} +
\nabla^2\rho$ with $2\nabla^2\rho$ in (\ref{P_Sch}) one obtains
the J\'anossy-Ziegler pressure \cite{JanZie63a,Har66a}. This is a
(gauge) freedom in case of neutral fluids but it can be important
when vorticity is not zero. In our thermodynamic derivation the
pressure is the primary constitutive quantity, it is determined
uniquely and the potentializability is the consequence of the
thermodynamic structure.
\end{rem}

\underline{Landau fluid}. The simplest globally concave and
isotropic entropy depends on the square of the density gradient.
Because of the similarity to the Ginzburg-Landau free energy
density, we will call Ginzburg-Landau fluid the material that is
defined by the following form of the entropy
\begin{equation}
    s\Lan(\rho,\nabla\rho,{\bf v}) =
        - \nu\Lan\frac{(\nabla\rho)^2}{2}
        - \frac{{\bf v}^2}{2}.
\label{Lf_s}\end{equation}

\noindent where $\nu\Lan$ is a constant coefficient. The
corresponding reversible pressure function is
\begin{equation}
{\bf P}\rev\Lan =
    - \nu\Lan \frac{\rho^2}{2}\left(\Delta\rho{\bf I}
    + \nabla^2\rho\right).
\label{P_L}\end{equation}

The potential is
\begin{equation}
    U\Lan = - \nu\Lan \frac{1}{2}\left(\rho \Delta \rho +
        \Delta\frac{\rho^2}{2}\right).
\label{U_L}\end{equation}

The entropy current can be also given, as
\begin{equation}
{\bf j}\Lan = -{\bf v}\cdot{\bf P}\rev\Lan
    - \nu\Lan \frac{\rho^2}{2}(\nabla\rho\nabla\cdot{\bf v}
       + \nabla\rho\cdot\nabla{\bf v}).
\label{js_L}\end{equation}

\underline{Alternative fluid}. The potential has the simplest form
if the entropy is written as
\begin{equation}
    s\Alt(\rho,\nabla\rho,{\bf v}) =
        - \nu\Alt\frac{(\nabla\rho)^2}{4\rho}
    - \frac{{\bf v}^2}{2}.
\label{Af_s}\end{equation}

\noindent where $\nu\Alt$ is a constant coefficient. In this case
the reversible pressure is
\begin{equation}
    {\bf P}\rev\Alt =
        -\frac{\nu\Alt}{4}\left(\rho(\Delta\rho{\bf I} + \nabla^2\rho)
            -\nabla\rho\circ\nabla\rho\right).
\label{P_A}\end{equation}

\noindent and the nonlocal term in the potential is simply
\begin{equation}
    U\Alt = - \frac{\nu\Alt}{2}\Delta\rho.
\label{U_A}\end{equation}

\noindent Finally, the entropy current can be written as
\begin{equation}
    {\bf j}\Alt = -{\bf v}\cdot{\bf P}\rev\Alt
       - \nu\Alt \frac{\rho}{4}(\nabla\rho\nabla\cdot{\bf v}
       + \nabla\rho\cdot\nabla{\bf v}).
\label{js_A}\end{equation}

\section{The origin of quantum potential}

An important property of the Schr\"odinger-Madelung fluid is that
if the motion of the fluid is vorticity free, $\nabla\times{\bf v}
={\bf 0}$, then the mass and momentum balances can be transformed
into and united in the Schr\"odinger equation. Hence the balance
of momentum (\ref{cond_mom}) can be derived from a Bernoulli
equation (in a given inertial reference frame). Defining a scalar
valued phase (velocity potential) by
$$
{\bf v} = \frac{\hbar}{m} \nabla S,
$$

\noindent the Bernoulli equation is obtained from the second part of
(\ref{balNew}) written as
\begin{equation}
\frac{\hbar}{m}\frac{\partial S}{\partial t} =
-\frac{{\bf v}^2}{2} + U\SchM.
\label{Bern_eq}\end{equation}

Then, introducing a single complex valued function $\psi := R e^{i
S}$ that unifies $R = \sqrt{\rho}$ and $S$, it is easy to find
that sum of (\ref{cond_mass}) multiplied by $i \hbar e^{i S}/(2R)$
and (\ref{Bern_eq}) multiplied by $m R e^{i S}$ form together the
Schr\"odinger equation for free particles
\begin{equation}
i \hbar \frac{\partial \psi}{\partial t} = -\frac{\hbar^2}{2 m}
\Delta\psi.
\label{Schro_eq}\end{equation}

Therefore, what we have accomplished is a kind of derivation of
the quantum mechanics in a very general framework. Moreover, we
have obtained some more, different but related fluid models. At
this point several questions can arise. What is the relation of
this thermodynamic derivation to other approaches? Are there
'quantized' solutions of the Landau or alternative fluids under
similar conditions that in quantum mechanics? Are there any
distinctive properties of the Schr\"odinger-Madelung model among
the other weakly nonlocal fluids?

First of all, to understand the results of Liu procedure from a
different point of view, let us observe that equation
(\ref{qpres_gen}) has an Euler-Lagrange form. Therefore, one can
derive it from a traditional Hamiltonian variational principle
with the Lagrangian density $\mathcal{ L }(\rho,\nabla\rho) =
-\rho s\static(\rho,\nabla\rho)$!

Moreover, in case of rotational free motion, the mass and momentum
balances can be calculated from a variational principle (see e.g.
\cite{Spi80a,Reg98a}). In our thermodynamic train of thought the
existence of a Hamiltonian variational principle was not a
starting point, but a consequence of the Second Law in a special
reversible case.

It is interesting to observe that the Schr\"odinger-Madelung model
is not distinguished from the other two fluids in producing static
"quantum" solutions. The static fluid dynamic equations give a
generalized eigenvalue problem
\begin{equation}
\nabla\cdot(\rho\partial_{\nabla\rho}s\static) -
    \partial_\rho(\rho s\static) = U = const.
\end{equation}

Because $U$ is given as a functional derivative of $\rho
s\static$, the existence of multiple solutions is connected to the
concavity properties of the entropy function. Among the previous
three fluids only the Landau fluid preserves the global concavity
of the entropy both in the density and in the gradient of the
density. One can calculate easily, that the second derivatives of
the entropies of the Schr\"odinger-Madelung fluid and of the
alternative fluid are positive semidefinite. The
Schr\"odinger-Madelung fluid has a rich structure of stationary
solutions in case of simple conservative force fields (the same as
for the Schr\"odinger equation). Since this property seems to be
connected to the semidefinite property of the entropy
\cite{VanFul04a}, the existence of similar "quantized solutions"
can be expected in the alternative fluid, too. This is an
interesting question both from a physical and from a mathematical
points of view. Evidently, a large set of different generalized
eigenvalue problems emerge.

Let us remark here that the concavity of the entropy (kinetic
potential) is also connected to the stability properties of the
equilibrium solutions of the fluid dynamic equations
\cite{VanFul04a}.

Finally, we turn our attention to the form of the static part of
the specific entropy function. Apart from a constant multiplier,
it can be written as $s\SchM(\rho,\nabla \rho) = \frac{(\nabla
\rho)^2}{\rho}$, which is the trace of the Fisher information
tensor of the probability density $\rho$ \cite{Bor98b}. Here we
will show shortly that this special form is a straightforward
consequence of some general properties of the theory and of the
entropy function $s\static(\rho,\nabla \rho)$. In the following
arguments we require some basic properties of point masses,
encoded in a continuum theory.

First of all, the entropy is {\em isotropic}. An isotropic
function $s$ of $\rho$ and $D\rho$ has the following property
\cite{PipRiv59a,Smi64a}
\begin{equation}
s\static(\rho,\nabla\rho) = \hat{s}(\rho,(\nabla\rho)^2).
\label{isotrop}\end{equation}

On the other hand, for independent particles one requires the
separability of the governing equations, therefore the {\em
additivity} of the entropy function in a definite way. We
postulate that, in case of independent particles, the governing
equation should be additively separable and, at the same time the
probability interpretation has to be preserved with considering
the entropy of the system of particles. This is exactly the
physical property why one requires the linearity of the
Schr\"odinger equation and any of its dissipative extensions
written in probability amplitude (wave function) variables.

For the sake of simplicity we restrict ourself for two particles.
The generalization to finite number of independent particles is
straightforward. For two distinct particles we can consider the
two-particle probability density $\rho({\bf x}_1,{\bf x}_2)$ which
is the product of the one particle probability densities
$\rho_1({\bf x}_1)$ and $\rho_2({\bf x}_2)$. Then the gradient of
$\rho$ is meant in two variables, too. Thus, we have
$(\nabla\rho)({\bf x}_1,{\bf x}_2) = (\rho_2({\bf
x}_2)\nabla_{{\bf x}_1}\rho_1({\bf x}_1),\rho_1({\bf
x}_1)\nabla_{{\bf x}_2}\rho_2({\bf x}_2))$ and - omitting the
variables ${\bf x}_1$ and ${\bf x}_2$ - $(\nabla\rho)^2 =
(\rho_2\nabla\rho_1)^2 + (\rho_1\nabla\rho_2)^2$. As a
consequence, in case of isotropic entropies the additivity
requirement can be written as:
\begin{equation}
\hat{s}(\rho_1\rho_2, (\rho_2\nabla\rho_1)^2 +
    (\rho_1\nabla\rho_2)^2) = \hat{s}(\rho_1,(\nabla\rho_1)^2) +
    \hat{s}(\rho_2,(\nabla\rho_2)^2).
\label{defext}\end{equation}

If $\hat{s}$ is continuously differentiable, then differentiating
the above equality by $(\nabla\rho_1)^2$ and $(\nabla\rho_2)^2$
respectively we have that
\begin{eqnarray*}
\rho_2^2 D_2 \hat{s}(\rho_1\rho_2, (\rho_2\nabla\rho_1)^2 +
(\rho_1\nabla\rho_2)^2) &=&
D_2 \hat{s}(\rho_1,(\nabla\rho_1)^2), \\
\rho_1^2 \partial_2\hat{s}(\rho_1\rho_2, (\rho_2\nabla\rho_1)^2 +
(\rho_1\nabla\rho_2)^2) &=&
D_2\hat{s}(\rho_2,(\nabla\rho_2)^2) .
\end{eqnarray*}

Here $D_2$ denotes the partial derivative of $\hat{s}$ by
its second argument. Therefore
\begin{equation*}
\rho^2 D_2\hat{s}(\rho,(\nabla\rho)^2) = \nu = const.,
\end{equation*}

\noindent from which it follows, that
\begin{equation}
\hat{s}(\rho, (\nabla \rho)^2) = \nu \frac{(\nabla\rho)^2}{\rho^2}
+ \tilde{s}(\rho),
\label{Fisform}\end{equation}

\noindent where $\tilde{s}$ is an arbitrary function (the local
part of the entropy). Repeating the above argument with the
derivatives by the first argument of $\hat{s}$, one finds that
\begin{equation*}
\rho \partial_1\tilde{s}(\rho) = k = const.
\end{equation*}

Consequently, $\tilde{s}(\rho) = k \ln \rho + s_0$, where $s_0$ is
an arbitrary constant. Therefore
\begin{equation}
\hat{s}(\rho, (\nabla \rho)^2) = \nu \frac{(\nabla\rho)^2}{\rho^2}
+ k \ln \rho + s_0. \label{EPIform}\end{equation}

The first term has the form of a Fisher information and the second
term has the form of a Shannon information measure. The solution
is unique with the above requirements.

Let us consider now, that our probability continuum should be a
theory of particles. Therefore we require the {\em mass-scale
invariance} of the entropy, that is the entropy density is a first
order homogeneous function of the density. In this case the
specific entropy is scale invariant, a zeroth order homogeneous
function of the density
\begin{equation}
s\static(\lambda \rho, \nabla(\lambda\rho)) =s\static(\rho,
\nabla\rho), \label{partcond}\end{equation}

\noindent for any real number $\lambda$. This a necessary
condition to have a particle interpretation of a continuum. In
this case the quantum potential is independent of the mass-scale
and thus the equations of free motion also have the same property.
One can see this clearly from the particle equation form of the
momentum balance (the second part of (\ref{balNew})). Considering
this condition the mass, the measure of the inertia of the
continuum acts uniformly, independently of the space coordinates.
With using specific entropy, that depends only on the specific
quantities (the density is the reciprocal value of the specific
volume) we required a volume-scale invariance of the corresponding
functions. The mass-scale invariance is something similar, it is a
kind of mass-extensivity.

Mass-scale invariance requires that $k=0$ in (\ref{EPIform})
excluding the logarithmic part.

It is interesting to note that a weakly nonlocal statistical
theory is developing that investigates the foundations of
equilibrium probability distributions and static nonlocal
extensions of fundamental equations of mathematical physics from
the point of view of Fisher information
\cite{PlaPla95a,Fri90a,FriAta02a}. However, the relationship
between Fisher information and thermodynamic entropy in
non-equilibrium situations is not clear and especially the
nonnegativity of the entropy production is questioned
\cite{FriAta99a,Net02a,Net03a}. In the light of the recent
investigations for the case of the Schr\"odinger equation the
entropy inequality is identically fulfilled.

\section{Conclusions}

In this paper we have shown that the Second Law of thermodynamics
restricts considerably the possible pressure functions of fluids
that are weakly nonlocal in density. Several different traditional
fluids were defined, where the non-equilibrium specific entropy is
additively quadratic in the velocity (\ref{tradf_s}), by some
simple possible forms of the nonlocal part of the entropy
function. In the conservative limit, when the entropy production
is zero, we have found that the entropy density is a Lagrangian of
the gradient dependent potentials, therefore, in case of vorticity
free motion it can be substituted for example into a
Seliger-Whitham-type variational principle
\cite{SelWhi68a,Spi80a}. However, in our thermodynamic approach,
the Euler-Lagrange form was a consequence of the Second Law in the
nondissipative limit. There was no need to refer to any
variational principle at all.

We have seen that the Schr\"odinger-Madelung fluid, the
hydrodynamic model of quantum mechanics plays a distinguished role
among the thermodynamically possible weakly nonlocal fluids. It is
the only model where the  nonlocal part of the entropy function is
isotropic, additive and mass-scale invariant. All these three
properties have a clear physical meaning. We need an additive
entropy to obtain independent equations for independent free
particles. Additivity requires isotropy. We need a mass-scale
invariant entropy to have a mass that measures the inertia of the
particle in our continuum theory. Only in this case will the
inertia of the ``fluid" be uniform and proportional to the force.

From a fluid mechanical point of view, the Schr\"odinger equation
appears as a complex formulation of the coupled Bernoulli equation
and the mass balance and has the great advantage of being linear.
On the other hand, we should emphasize that our derivation of the
hydrodynamic model of quantum mechanics was completely independent
of the Schr\"odinger equation, and quantum mechanics appeared only
as a special fluid with remarkable properties. There are several
examples of the applicability of generalized fluid models.
Generalized Schr\"odinger-type equations appear as structure
forming equations in several fields without any connection to
quantum mechanics \cite{AraKra02a}. The possibility to transform
generalized weakly nonlocal fluids into a well known linear
complex form is an important advantage in several investigations
as, e.g., in cosmology \cite{WidKai93a,SzaKai02m,Col02m}.

Let us give some remarks on other derivations of the basic
equation of quantum mechanics. As it is well known, Schr\"odinger
himself did not derive the equation, in a strict sense, his
suggestion is based on analogies and is justified by its
consequences. The de Broglie-Bohm form or the hydrodynamic model
of Madelung all start from the Schr\"odinger equation, so they are
not derivations only different points of view given for
interpretational purposes (nevertheless important and thought
provoking ones).

The stochastic model \cite{Fen52a,Nel66a,PenCet96b,FriHau03a}
provides a kind of derivation. Here one assumes a background
random field, the active role of "vacuum fluctuations" as an
origin of the quantum potential, but one should not forget about
the special assumed properties. The introduced stochastic velocity
has a particular form, being proportional to the gradient of
$\rho$, which form is motivated by the analogy (!) with diffusion.
Here nonlocality is disguised as a kind of velocity, hence
requires a special structure and is similar - results in - the
mass-scale invariance of the entropy.

Nonlocal kinetic theories (e.g. that of Kaniadakis \cite{Kan02a})
all assume a definite microscopic background.

In the previous section we have mentioned other approaches based
on the Fisher information measure. These start from a variational
principle and postulate the form of quantum term (as minimum
Fisher information e.g., \cite{Reg98a}) and interpret it with
information theoretical ideas based on the measurement process
\cite{Fri98b}. Such an approach is thus less a derivation but
rather a different interpretation (nevertheless important and
thought provoking).

The only remarkable exception is the approach of Hall and
Reginatto \cite{HalReg02a,HalReg02a1} who give arguments on the
form of the entropy (in their formulation it is an additional term
in the Lagrangian due to fluctuations) with a reasoning that is
similar to ours. They require additivity and isotropy
(implicitly), but instead of mass-scale invariance they use a
different principle what they call "exact uncertainty".
Nevertheless with these requirements they arrive at the same
Fisherian form of the quantum part of the entropy function. On the
other hand they start from a variational approach which is a
consequence of the Second Law in our considerations.

We have to emphasize again that our derivation here is based on
general principles (Second Law) and requirements (additivity) and
therefore is independent on any kind of interpretational issues of
quantum mechanics. It contains a minimal set of assumptions that
one can gain on a phenomenological level about one component fluid
systems. It is surprising that we have reached almost everything.
The only thing in the equations that one should determine from
microscopic considerations (from experiments) is the value of the
Planck constant.

Our approach is similar that of Jaynes to equilibrium statistical
physics both on the dynamic and on the static level \cite{Jay57a}.
On the dynamic level we used the entropy inequality - a part of
the Second Law - and the basic balances as constraints to extract
the maximum amount of information regarding the structure of a
definite physical system. We have derived serious restrictions on
the constitutive functions. Moreover, the reasoning is predictive
in the Jaynesian sense. E.g. in the dynamic case requiring
nonnegative entropy production one can give definite predictions
on the structure of nonlocal equations of multicomponent fluids,
too. Considering phase separation as a constraint, we can
construct promising new models for granular and porous media
\cite{Van04a,Van04a1}.

Moreover, regarding the static part, our reasoning is completely
analogous to the Jaynesian, phenomenological approach of
statistical physics. Jaynes's arguments are based on the
uniqueness of Shannon's information measure with some expected
properties that are common with the required properties of the
entropy function (extensivity, additivity). However, Shannon's
proof was related to a discrete probability space and exploited
that the entropy is a composite (local) function. Here, we can
regard our approach as a kind of nonlocal extension of the
Jaynes-Shannon argumentation. The form (\ref{EPIform}) of the
entropy is unique with the given requirements, therefore one can
use it as a starting point. The emergent structure is independent
of the microscopic background. Every reasonable microscopic
approach (kinetic or stochastic) corresponds to the formulated
general requirements.

Moreover, in (\ref{EPIform}) one can recognize the central formula
of the Extreme Physical Information (EPI) principle of Frieden
\cite{Fri98b} and our argumentation gives a kind of foundation of
the principle (uniqueness) and also some limitations (conditions
of validity as mass-scale invariance) and a different
interpretation (there is no need of the measurement-information
arguments, objectivity of the theory is reconstructed). Here the
predictivity of the approach is even more evident as one can see
from the increasing number of applications of  statistical physics
based on Fisher information.

We think that the recent point of view and results can be
generalized far beyond quantum mechanics, can be used as a general
approach to non-equilibrium and weakly nonlocal statistical physics.

\section{Acknowledgements}

This research was supported by OTKA T034715 and T034603. The authors
wish to say thanks to Tam\'as Matolcsi from whom they have learned that
quantum mechanics is a closed but unfinished theory, and thank J.\
Verh\'as, S.\ Katz and K.\ Ol\'ah for the discussions regarding several
different topics of physics including the foundations of quantum
mechanics and thermodynamics.

\end{document}